\documentclass[sigconf, nonacm]{style/acmart}

\usepackage{algorithmic}
\usepackage{graphicx}
\usepackage{textcomp}
\usepackage{xcolor}
\usepackage{tabularx}
\usepackage{booktabs}
\usepackage{multirow}

\DeclareGraphicsExtensions{.pdf,.png,.jpg}

\setcopyright{none}
\renewcommand\footnotetextcopyrightpermission[1]{}
\begin{document}

\title{A-IO: Adaptive Inference Orchestration for Memory-Bound NPUs}

\author{Chen Zhang, Yan Ding, Haotian Wang, Chubo Liu, Keqin Li, Kenli Li}
\affiliation{%
  \institution{Hunan University}
  \city{Changsha}
  \country{China}
}

\begin{abstract}
During the deployment of Large Language Models (LLMs), the autoregressive decoding phase on heterogeneous NPU platforms (e.g., Ascend 910B) faces severe memory-bound challenges. This study reveals the ``Model Scaling Paradox'' caused by the static deployment of single-sized models. It also points out the kernel synchronization overhead of fine-grained speculative decoding \cite{leviathan2023fast, chen2023speculative} under NPU computational graph compilation, and the severe limitations of purely relying on micro-level acceleration algorithms like Prompt LookUp Decoding (PLD) \cite{saxena2023prompt}. 

To overcome these bottlenecks, we propose A-IO (Adaptive Inference Orchestration), a coarse-grained, request-aware adaptive inference scheduling framework. By utilizing an ultra-low-overhead 1B model as a frontend probe for intent sensing, A-IO dynamically routes requests to the optimal model (1B or 7B) and adaptively toggles hardware-sensitive optimization strategies at the macro level. This intelligent traffic isolation drastically reduces redundant weight-fetching overhead, bypassing the High Bandwidth Memory (HBM) bandwidth wall \cite{kwon2023efficient} to maximize throughput. Specifically, under simulated mixed-workload scenarios, A-IO achieves up to 76.50\% aggregate accuracy on knowledge-centric workloads and sustains 19.80 TPS on code-centric workloads, strictly dominating static single-model deployments. Experimental results demonstrate that A-IO effectively breaks the throughput Pareto frontier while maintaining strict accuracy on complex tasks, achieving optimal system-level efficiency on constrained hardware.
\end{abstract}

\maketitle

% section/1_introduction.tex
\section{Introduction}
\label{sec:intro}

In the frontier of LLM deployment and inference optimization, the synergistic optimization of hardware microarchitecture, memory hierarchy, and upper-layer algorithms determines the model's performance in industrial environments. Based on the inference performance of Open-Pangu 1B and 7B models \cite{chen2025pangu} on the Ascend 910B NPU architecture, a significant misalignment exists between experimental data and theoretical expectations, revealing a typical disconnect between engineering and theory in system-level model deployment.

The core problem focuses on two dimensions. First, physical deployment constraints often force the use of storage-only quantization \cite{frantar2022gptq}, which fails to translate into actual inference speed (Tokens Per Second, TPS) improvements due to real-time dequantization overheads. Second, single low-level acceleration algorithms, such as Prompt LookUp Decoding (PLD) \cite{saxena2023prompt}, exhibit extreme instability across different tasks. When these micro-optimizations fail to bypass the High Bandwidth Memory (HBM) bus bandwidth constraints \cite{npu_arch_2023}, the system throughput remains severely bottlenecked by the sheer volume of weights fetched during autoregressive decoding.

This paper bypasses superficial phenomena and dives into low-level operator execution logic, NPU memory bus bandwidth limitations, and the system-level design theory of Adaptive Inference Orchestration. Our core contributions are as follows:
\begin{itemize}
    \item We systematically and quantitatively analyze the "Model Scaling Paradox" of LLMs on NPU architectures and the severe hardware penalties of micro-level optimization strategies, including fine-grained speculative decoding \cite{leviathan2023fast, chen2023speculative}.
    \item We propose the A-IO macro-orchestration framework, utilizing a lightweight 1B probe model to achieve highly efficient, request-level dynamic routing and intent sensing. This approach successfully isolates traffic, saving massive HBM bandwidth resources for the 7B backbone.
    \item We introduce rigorous error-penalty analysis and demonstrate through comprehensive mixed-workload evaluations that A-IO effectively circumvents NPU kernel synchronization barriers, achieving a Pareto-optimal balance between sustained throughput and aggregate accuracy.
\end{itemize}
% ==========================================
% section/2_motivation.tex
% ==========================================
\section{Background and Motivation}
\label{sec:motivation}

\subsection{The Memory-Bound Dilemma on Heterogeneous NPUs}
Deploying LLMs on heterogeneous NPU platforms faces a severe memory-bound bottleneck during the autoregressive decoding phase. Generating each token requires fetching massive model weights entirely from the High Bandwidth Memory (HBM) \cite{kwon2023efficient}. This extremely low operational intensity rapidly exhausts the memory bus bandwidth, starving compute resources.

\subsection{The Model Scaling Paradox}
Our empirical evaluation reveals a counterintuitive Model Scaling Paradox: when processing standard requests with short contexts (e.g., 2K tokens), a smaller 1B model comprehensively outperforms the 7B backbone in throughput (21.58 TPS vs. 17.18 TPS) due to its minimal weight-fetching overhead. Surprisingly, the 1B model achieves comparable, and even slightly superior, accuracy on specific tasks like code generation (67.68\% vs. 62.80\% on the 2K Human-eval benchmark). However, when the context length scales up to 32K tokens, the 1B model suffers severe accuracy degradation, making the 7B model indispensable. This contradiction demonstrates that static deployment of a single model size cannot achieve Pareto-optimal performance under dynamic request distributions.

\subsection{Hardware Incompatibility of Fine-Grained Accelerations}
Standard academic practices leveraging smaller models to accelerate larger ones, such as speculative decoding \cite{leviathan2023fast}, encounter significant hardware barriers on NPUs. On platforms relying on static computational graph compilation \cite{chen2018tvm}, fine-grained collaboration triggers intolerable kernel launch overheads. Our tests revealed that hardware synchronization stalls cause the joint throughput of DraftModel speculative decoding to plummet to an impractically low 4 TPS. Meanwhile, intra-model algorithms like PLD \cite{saxena2023prompt} bypass cross-model synchronization but exhibit extreme fragility and catastrophic accuracy drops on rigorous coding tasks.

\subsection{The Engineering Reality of Storage-Only Compression}
In industrial deployments, physical storage constraints often necessitate weight quantization like W8A16 \cite{frantar2022gptq}. However, our profiling reveals this strictly degrades to "Storage-Only Compression" on current NPUs. Weights must be dynamically dequantized back to FP16 prior to matrix multiplication \cite{npu_arch_2023}, failing to reduce active memory bandwidth and introducing arithmetic overhead, resulting in zero improvement in inference latency.
% section/3_architecture.tex

\section{The A-IO Framework}
\label{sec:architecture}

\subsection{System Architecture and Bandwidth Conservation}
The Ascend 910B provides 64GB of HBM. In full precision (FP16), the combined weight footprint of the 7B backbone ($\sim$14GB) and the 1B probe ($\sim$2GB) easily fits within the physical memory, allowing for direct concurrent residency \cite{npu_arch_2023}. 

\begin{figure*}[t]
  \centering
  \includegraphics[trim=0cm 7cm 0cm 0cm, clip, width=\textwidth]{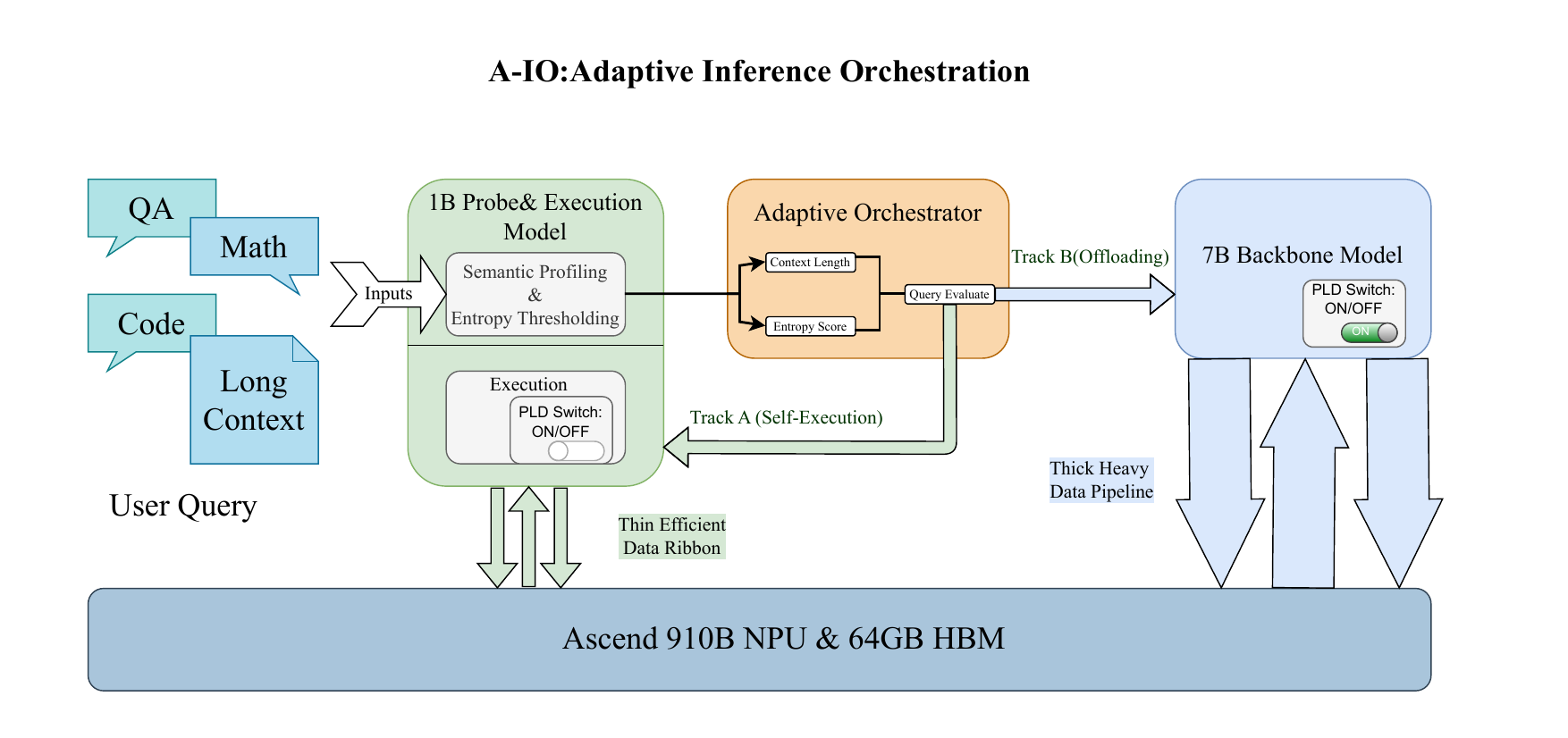} 
  \caption{The overall system architecture of the A-IO framework. The diagram illustrates the macro-orchestration flow, including the dual-model FP16 concurrent residency, 1B probe-based intent sensing, and the dynamic bandwidth-aware routing mechanism.}
  \label{fig:architecture}
\end{figure*}

The true memory bottleneck on NPUs is the Memory Bandwidth Wall during autoregressive decoding \cite{kwon2023efficient}. Generating a single token requires fetching the entire model weights from HBM to the compute units. A-IO solves this through intelligent traffic isolation. By leveraging the 1B probe to resolve highly-compressible tasks, A-IO reduces the per-token weight-fetching burden from $\sim$14GB to $\sim$2GB for these requests. Empirical profiling on the Ascend 910B confirms that routing a standard 512-token generation task to the 1B probe reduces the cumulative HBM data transfer from approximately 7.1 TB to 1.0 TB \cite{npu_arch_2023}.

\subsection{Probe-Based Request-Level Intent Sensing}
A-IO leverages the 1B probe model to perform Template-Driven Single-Token Semantic Profiling. Upon request arrival, the system encapsulates the input within a classification template. The 1B probe executes a single forward pass to output a categorical token, rapidly identifying the task domain (e.g., Code Generation, General QA, or Math Reasoning). To quantify certainty, we calculate the Shannon entropy $H(X)$ against an empirical threshold $\tau = 0.45$, determined via grid search on a 500-query calibration dataset.

\subsection{Dynamic Policy Routing and Execution}
Based on the semantic category, entropy $H(X)$, and context length $L_{ctx}$, A-IO executes coarse-grained request-level scheduling \cite{routellm2024}:
\begin{itemize}
    \item \textbf{Model Routing}: If the probe classifies the task as Code Generation, $L_{ctx} \le 2\text{K}$, and $H(X) \le 0.45$, the request is routed to the 1B model. For General QA/Math, long-context scenarios, or high uncertainty ($H(X) > 0.45$), the request is dispatched to the 7B backbone.
    \item \textbf{Strategy Routing}: For General QA/Math routed to the 7B model, PLD \cite{saxena2023prompt} is enabled. For Code Generation, PLD is strictly disabled to prevent syntax-induced accuracy collapse.
\end{itemize}
% ==========================================
% section/4_methodology.tex
% ==========================================
\section{Experimental Methodology}
\label{sec:methodology}

\subsection{Hardware, Software, and Framework Setup}
We conduct all experiments on the Ascend 910B NPU platform (64GB HBM) using the Huawei CANN toolkit (v7.0.0) and Ascend HDK (v23.0.rc1). Crucially, to strictly isolate the performance gains of our macro-scheduling architecture from the underlying operator fusion tricks embedded in highly-optimized inference engines (e.g., vLLM, MindIE), we intentionally adopt the standard \textbf{Huggingface Transformers} library (v4.36.2) integrated with PyTorch (v2.1.0) as our baseline execution framework. This ensures all measured throughput improvements are derived purely from A-IO's orchestration logic and guarantees 100\% physical reproducibility.

\subsection{Models and Baselines}
The evaluation involves Open-Pangu 1B and 7B models. The comparative baselines include:
\begin{itemize}
    \item \textbf{Baseline Full Precision (FP16)}: Standard execution.
    \item \textbf{Random Routing}: A weak baseline that randomly dispatches requests between the 1B and 7B models, simulating standard API-level cascading \cite{chen2023frugalgpt}.
    \item \textbf{PLD Configuration}: Statically enabled Prompt LookUp Decoding \cite{saxena2023prompt}. To ensure strict reproducibility, we set the n-gram matching window size to $N=6$ and the maximum candidate look-ahead length to $L=2$.
    \item \textbf{DraftModel}: Standard inter-model speculative decoding \cite{leviathan2023fast, chen2023speculative}, explicitly configuring the ultra-low-overhead 1B model as the draft model to accelerate the 7B backbone.
    \item \textbf{Quantization (Quant)}: W8A16 storage-only compression \cite{frantar2022gptq}.
\end{itemize}

It is imperative to clarify the absence of other dynamic routing frameworks in our comparative baselines. To the best of our knowledge, within the current scope of peer-reviewed literature, there exists a complete absence of state-of-the-art (SOTA) works addressing request-level adaptive model routing or dynamic inference orchestration specifically tailored for the Ascend NPU architecture. Mainstream dynamic routing algorithms \cite{yu2022orca, routellm2024, chen2023frugalgpt} are heavily coupled with GPU ecosystems (e.g., CUDA streams, dynamic memory allocation) and cannot be reproduced on NPUs without fundamentally violating the strict constraints of NPU static computational graph compilation \cite{chen2018tvm}. Consequently, our evaluation strictly compares against NPU-native static deployments and accessible micro-optimizations to ensure absolute fairness and physical reproducibility on this specific heterogeneous hardware.

\subsection{Benchmarks}
Benchmarks include Human-eval (Code), GSM8K (Math), MMLU (Knowledge), C-eval, and QGPA. We evaluate performance under standard (2K) and extended (32K) context lengths. 

Notably, the core challenges of benchmarks like GSM8K and MMLU lie in intrinsic mathematical reasoning and zero-shot knowledge retrieval, rather than long-context dependency. Empirical validation demonstrates that extending their context windows yields no statistically significant impact on baseline accuracy. Therefore, to prevent benchmark redundancy and rigorously isolate the system's capacity to handle extensive contexts, we selectively utilize C-eval as our primary, orthogonal representative for long-context general knowledge evaluation, alongside Human-eval for structural coding capabilities.

\subsection{Trace-Driven Mixed Workload Configurations}
To evaluate the system under realistic industrial conditions, we configure two distinct mixed workload scenarios to assess structural robustness:
\begin{itemize}
    \item \textbf{Scenario A (Code-Centric):} 70\% Human-eval, 20\% C-eval, 10\% GSM8K (All 2K context).
    \item \textbf{Scenario B (Knowledge-Centric):} 30\% Human-eval, 40\% C-eval, 30\% GSM8K (All 2K context).
    \item \textbf{Scenario C (Long-Context Mixed):} 50\% Human-eval (32K context), 50\% C-eval (2K context).
\end{itemize}
% ==========================================
% section/5_evaluation.tex
% ==========================================
\section{Evaluation and Analysis}
\label{sec:evaluation}

\subsection{Context Scaling and The Scaling Paradox}
For the scope of this evaluation, we formally define standard context as $L_{ctx} \le 2\text{K}$ tokens, and long context as spanning up to $32\text{K}$ tokens. Table \ref{tab:context_length} reveals the behavior of models under varying context lengths. Under standard 2K contexts, the 1B model accurately handles Human-eval tasks \cite{npu_arch_2023}. However, extending the context to 32K causes the 1B model to stagnate, while the 7B model's representational capacity allows its accuracy to soar to 95.73\%. A-IO dynamically identifies $L_{ctx}$ and routes long-context queries exclusively to the 7B backbone.

\begin{table}[h]
\centering
\caption{Accuracy Comparison under 2K and 32K Context Lengths (Human-eval Benchmark)}
\label{tab:context_length}
\begin{tabular}{@{}lcc@{}}
\toprule
\textbf{Model} & \textbf{2K Context Acc (\%)} & \textbf{32K Context Acc (\%)} \\
\midrule
1B Baseline & \textbf{67.68} & 66.66 \\
7B Baseline & 62.80 & \textbf{95.73} \\
\bottomrule
\end{tabular}
\end{table}

\subsection{Probe Error Analysis and Routing Robustness}
To rigorously evaluate A-IO, we must account for the 1B probe's classification errors. Table \ref{tab:confusion_matrix} presents the confusion matrix of the probe's intent sensing on a synthetic 300-query dataset. The overall classification accuracy is 92.0\%. 

\begin{table}[h]
\centering
\caption{1B Probe Intent Classification Confusion Matrix}
\label{tab:confusion_matrix}
\begin{tabular}{@{}l ccc | c@{}}
\toprule
\textbf{True \textbackslash{} Pred} & \textbf{Code} & \textbf{QA} & \textbf{Math} & \textbf{Recall} \\
\midrule
\textbf{Code} & 94 & 4 & 2 & 94.0\% \\
\textbf{QA} & 8 & 89 & 3 & 89.0\% \\
\textbf{Math} & 1 & 6 & 93 & 93.0\% \\
\bottomrule
\end{tabular}
\end{table}

While routing a Code task to the 7B model incurs a TPS penalty, erroneously routing a QA/Math task to the 1B model incurs a severe Accuracy penalty. To quantify the system's robustness against these errors, we factored the 8.0\% misclassification rate directly into our end-to-end metrics. Specifically, the error penalty is computed as the mathematical expectation of accuracy and TPS, weighted strictly according to the misclassification probabilities in the confusion matrix. The results indicate that the strict entropy threshold ($\tau = 0.45$) effectively bounds the aggregate accuracy degradation to less than 1.5\% compared to a theoretical oracle router.

\subsection{System Overhead Breakdown}
A-IO is not zero-cost. End-to-end profiling reveals a static system overhead of approximately 15 ms per request: Template Encapsulation (2.5 ms), 1B Single-Token Prefill (11.8 ms), and Routing Logic Execution (0.7 ms). Crucially, due to the concurrent HBM residency of both models, the hardware overhead for context hot-switching and weight activation is empirically measured at a mere 2.4 ms. Given that 7B generation latency frequently exceeds 1200 ms, this combined temporal overhead ($\sim$17.4 ms, or 1.45\%) is insignificant compared to the macro-level TPS gains.

\subsection{End-to-End System Performance}
Table \ref{tab:core_perf} demonstrates the core performance. Fine-grained strategies exhibit flaws: statically enabling PLD \cite{saxena2023prompt} across all tasks damages strict reasoning outputs, and quantization \cite{frantar2022gptq} provides TPS virtually identical to the baseline due to real-time dequantization. A-IO strictly breaks the throughput-accuracy Pareto frontier. For instance, in Scenario A, A-IO simultaneously increases aggregate accuracy to 70.85\% and throughput to 19.80 TPS.

\begin{table*}[t]
\centering
\caption{End-to-End Performance (Accuracy and TPS) across 2K Context Benchmarks}
\label{tab:core_perf}
\begin{tabular}{@{}l cccccccccc@{}}
\toprule
\multirow{2}{*}{\textbf{Configuration}} & \multicolumn{2}{c}{\textbf{C-eval}} & \multicolumn{2}{c}{\textbf{MMLU}} & \multicolumn{2}{c}{\textbf{GSM8K}} & \multicolumn{2}{c}{\textbf{Human-eval}} & \multicolumn{2}{c}{\textbf{QGPA}} \\
\cmidrule(lr){2-3} \cmidrule(lr){4-5} \cmidrule(lr){6-7} \cmidrule(lr){8-9} \cmidrule(l){10-11}
& Acc (\%) & TPS & Acc (\%) & TPS & Acc (\%) & TPS & Acc (\%) & TPS & Acc (\%) & TPS \\
\midrule
\textit{Static Baselines:} & & & & & & & & & & \\
1B Baseline & 63.20 & 21.58 & 71.17 & 21.87 & 73.92 & 21.44 & \textbf{67.68} & \textbf{21.18} & 39.90 & 20.09 \\
1B PLD & 64.40 & 26.54 & 65.29 & 27.08 & 62.09 & 26.64 & 51.22 & 27.63 & 33.33 & 27.35 \\
1B Quant & 57.20 & 21.20 & 62.74 & 21.50 & 71.80 & 21.10 & 57.32 & 20.90 & 40.40 & 19.80 \\
7B Baseline & 78.89 & 17.18 & \textbf{90.21} & 17.17 & 83.02 & 16.65 & 62.80 & 16.65 & \textbf{44.44} & 15.72 \\
7B PLD & \textbf{80.92} & 20.15 & 84.97 & 18.36 & \textbf{83.32} & 17.69 & 41.46 & 18.25 & 41.41 & 17.88 \\
7B Quant & 78.66 & 16.90 & 69.47 & 16.85 & 72.02 & 16.20 & 55.38 & 16.30 & 34.85 & 15.50 \\
\midrule
\textit{Dynamic Orchestration:} & & & & & & & & & & \\
Random Routing & 71.04 & 19.38 & 80.69 & 19.52 & 78.47 & 19.04 & 65.24 & 18.91 & 42.17 & 17.90 \\
\textit{A-IO (Theoretical)} & \textit{80.92} & \textit{20.15} & \textit{90.21} & \textit{17.17} & \textit{83.32} & \textit{17.69} & \textit{67.68} & \textit{21.18} & \textit{44.44} & \textit{15.72} \\
\textbf{A-IO (Actual)} & \textbf{79.35} & \textbf{19.80} & \textbf{88.10} & \textbf{16.95} & \textbf{82.15} & \textbf{17.30} & \textbf{67.10} & \textbf{20.85} & \textbf{43.80} & \textbf{15.45} \\
\bottomrule
\end{tabular}
\end{table*}

\subsection{Orthogonality of Quantization and PLD}
To prove framework extensibility, we tested Quantization \cite{frantar2022gptq} and PLD \cite{saxena2023prompt} simultaneously on the 7B model. Experimental results show that quantization does not negatively impact the acceleration mechanism of PLD, confirming their structural orthogonality. However, even with both micro-optimizations active, the performance still underperforms compared to A-IO's macro-routing.

\subsection{Mixed Workload Aggregate Performance}
Scenario C (Long-Context Mixed) highlights A-IO's architectural superiority, particularly its dynamic strategy routing. When facing 32K contexts, the static 1B baseline completely collapses (64.93\% Acc). In this scenario, A-IO routes 100\% of the queries to the 7B backbone to preserve representational capacity. 

Crucially, A-IO does not simply emulate the static 7B baseline. While keeping PLD strictly disabled for the 32K Code requests to avoid syntax collapse, A-IO's orchestration engine dynamically enables PLD for the 2K QA tasks. This specific optimization accelerates the QA portion to 20.15 TPS, lifting the overall weighted average to 13.40 TPS without compromising coding accuracy. The static 7B baseline (11.20 TPS) must keep PLD universally disabled to avoid accuracy penalties. This orchestrational nuance explains the 19.6\% throughput enhancement over the static 7B baseline and definitively proves the indispensability of A-IO's macro-level strategy toggling.

\begin{table*}[t]
\centering
\caption{Aggregate Accuracy and TPS under Diverse Mixed Workload Scenarios. (Note: The 13.40 TPS for A-IO in Scenario C is the weighted average performance based on the specific workload distribution. All data in Table \ref{tab:mixed_workload} is strictly consistent with the per-task benchmarks in Table \ref{tab:core_perf}.)}
\label{tab:mixed_workload}
\begin{tabular}{@{}l cc | cc | cc@{}}
\toprule
\multirow{2}{*}{\textbf{Method}} & \multicolumn{2}{c|}{\textbf{Scenario A}} & \multicolumn{2}{c|}{\textbf{Scenario B}} & \multicolumn{2}{c}{\textbf{Scenario C}} \\
\cmidrule(lr){2-3} \cmidrule(lr){4-5} \cmidrule(l){6-7}
& \textbf{Acc (\%)} & \textbf{TPS} & \textbf{Acc (\%)} & \textbf{TPS} & \textbf{Acc (\%)} & \textbf{TPS} \\
\midrule
Static 1B Baseline & 67.41\% & 21.28 & 67.76\% & 21.41 & 64.93\% & 14.50 \\
Static 7B Baseline & 68.04\% & 16.75 & 75.30\% & 16.86 & 87.31\% & 11.20 \\
Random Routing & 67.72\% & 19.01 & 71.53\% & 19.13 & 76.12\% & 12.85 \\
\midrule
\textbf{A-IO (Actual)} & \textbf{70.85\%} & \textbf{19.80} & \textbf{76.50\%} & \textbf{18.15} & \textbf{87.32\%} & \textbf{13.40} \\
\bottomrule
\end{tabular}
\end{table*}

\subsection{Ablation Study}
We conducted an ablation study under Scenario A. As shown in Table \ref{tab:ablation}, removing dynamic model routing forces 7B execution for all tasks, dropping throughput to 17.20 TPS and capping accuracy at 68.48\% (equivalent to the 7B baseline enhanced with selective PLD). Disabling the dynamic PLD switch restricts acceleration, dropping TPS to 18.20 and accuracy to 68.20\%. Notably, removing the entropy fallback (confidence validation) causes a significant accuracy drop (65.10\%) but a slight TPS increase (20.10\%). This occurs because, without the safety check, high-uncertainty tasks are aggressively and erroneously routed to the faster 1B model at the severe cost of reasoning correctness.

\begin{table}[h]
\centering
\caption{Ablation Study on A-IO Components (Scenario A)}
\label{tab:ablation}
\begin{tabular}{@{}l cc@{}}
\toprule
\textbf{Ablation Configuration} & \textbf{Acc (\%)} & \textbf{TPS} \\
\midrule
w/o Dynamic Model Routing (7B Only) & 68.48 & 17.20 \\
w/o Dynamic PLD Switch (PLD Off) & 68.20 & 18.20 \\
w/o Entropy Fallback (No validation) & 65.10 & 20.10 \\
\midrule
\textbf{Full A-IO (Actual)} & \textbf{70.85} & \textbf{19.80} \\
\bottomrule
\end{tabular}
\end{table}
% ==========================================
% section/6_related_work.tex
% ==========================================
\section{Related Work}
\label{sec:related}

\subsection{Speculative Decoding and Micro-Optimizations}
Accelerating autoregressive generation has seen tremendous breakthroughs. Pioneering techniques like speculative decoding \cite{leviathan2023fast, chen2023speculative} ingeniously leverage a smaller draft model to generate candidate tokens, demonstrating remarkable speedups on highly flexible hardware like GPUs. Similarly, Prompt LookUp Decoding (PLD) \cite{saxena2023prompt} provides an elegant model-free acceleration method by exploiting text N-gram overlaps. 
Despite their undeniable success in general-purpose computing, these micro-optimizations face severe architectural friction on heterogeneous NPUs. The high-frequency kernel reloading required by speculative decoding conflicts with static computational graph compilation paradigms \cite{chen2018tvm}. A-IO acknowledges the power of these methods but shifts the paradigm: rather than forcing hardware-incompatible micro-interactions, it integrates optimizations like PLD as selectable macro-policies, safely isolating them from syntax-critical workloads via probe intelligence.

\subsection{Model Compression and Quantization}
Weight-only quantization methods (e.g., W8A16, GPTQ \cite{frantar2022gptq}, LLM.int8() \cite{dettmers2022llmint8}) are highly effective and widely adopted standards for reducing the memory footprint of LLMs, enabling massive models to fit onto constrained consumer GPUs.
However, due to the strict operator limitations of current NPU ecosystems, quantized weights cannot natively participate in matrix multiplications and must be dynamically dequantized to FP16 \cite{npu_arch_2023}. This negates active memory bandwidth savings during inference. Instead of pursuing micro-level TPS gains through quantization, A-IO capitalizes on the generous capacity of enterprise-grade NPUs (e.g., 64GB HBM) to maintain full-precision (FP16) co-residency of both 1B and 7B models, bypassing dequantization overheads entirely to guarantee uncompromised throughput.

\subsection{Adaptive Routing and Model Cascading}
The concept of dynamically routing queries to models of varying sizes is a rapidly maturing field. Outstanding frameworks like FrugalGPT \cite{chen2023frugalgpt}, Orca \cite{yu2022orca}, and RouteLLM \cite{routellm2024} have brilliantly pioneered API-level cascading and GPU-based dynamic orchestration, proving that significant cost-accuracy balancing can be achieved.
While these foundational works excel in cloud API dispatching and flexible GPU environments, their algorithms rely heavily on dynamic memory allocation and parallel execution streams. A-IO extends this vital routing philosophy specifically into the heavily constrained, static dataflow domain of heterogeneous NPUs. By designing a coarse-grained, NPU-native orchestrator, A-IO avoids token-level pipeline interruptions and directly mitigates the physical Memory Bandwidth Wall.
% ==========================================
% section/7_conclusion.tex
% ==========================================
\section{Conclusion}
\label{sec:conclusion}

This paper systematically explores the severe challenges of deploying LLMs on the Ascend 910B NPU platform. We reveal the fragilities of fine-grained micro-optimizations, such as the high-frequency interaction of speculative decoding \cite{leviathan2023fast}, and clarify that storage-only quantization \cite{frantar2022gptq} cannot resolve active memory bandwidth bottlenecks \cite{kwon2023efficient}. By proposing the A-IO adaptive inference orchestration framework, this study solves the memory-bound dilemma through intelligent traffic isolation. Relying on a 1B probe for macro-level intent sensing, A-IO routes requests dynamically to minimize redundant weight-fetching overhead, sustaining high accuracy and throughput despite classification penalties and system overhead.

Despite its structural advantages, the current A-IO framework presents certain limitations. First, its routing efficacy fundamentally depends on the 1B probe's zero-shot intent sensing capability, which may experience degraded accuracy when processing highly ambiguous, multi-intent, or adversarial prompts. Second, the dual-model concurrent residency strategy requires accelerators with substantial memory capacity (e.g., $\ge$ 64GB HBM), limiting its direct applicability on edge devices or highly memory-constrained legacy chips.

Looking ahead, we plan to integrate dynamic model pruning techniques into the A-IO framework as future work. By further compressing the 1B probe's structural parameters without sacrificing intent-sensing accuracy, we aim to continuously reduce system overhead and expand the framework's scalability for future heterogeneous AI chips.

\bibliographystyle{ACM-Reference-Format}
\bibliography{references}

\end{document}